\documentclass[%
reprint,
superscriptaddress,
amsmath,amssymb,
aps,
]{revtex4-1}

\usepackage[usenames]{color}
\usepackage{colortbl}

\usepackage{graphicx}
\usepackage{dcolumn}
\usepackage{bm}

\begin{document}


\title{
Optical orientation of spins in GaAs:Mn/AlGaAs quantum wells via impurity-to-band excitation
}

\author{P.V.~Petrov}
 \email{pavel.petrov@gmail.com}
\affiliation{Ioffe Institute, St. Petersburg, Russian Federation}

\author{I.A.~Kokurin}
\affiliation{Ioffe Institute, St. Petersburg, Russian Federation}
\affiliation{Institute of Physics and Chemistry, Mordovia State
University, Saransk, Russian Federation}

\author{Yu.L.~Ivanov}
\author{N.S.~Averkiev}
\affiliation{Ioffe Institute, St. Petersburg, Russian Federation}

\author{R.P.~Campion}
\author{B.L.~Gallagher}
\affiliation{School of Physics and Astronomy, University of Nottingham, Nottingham, United Kingdom}

\author{P.M.~Koenraad}
\author{A.Yu.~Silov}
\email{A.Y.Silov@tue.nl}
\affiliation{Department of Applied Physics, Eindhoven University of Technology, The Netherlands}

\date{\today}

\begin{abstract}
The paper reports optical orientation experiments performed in
the narrow GaAs/AlGaAs quantum wells doped with Mn. We experimentally demonstrate
a control over the spin polarization by means of the optical orientation via
the impurity-to-band excitation and observe a sign inversion of
the luminescence polarization depending on the pump power.
The g-factor of a hole localized on the Mn acceptor in the
quantum well was also found to be considerably modified from its
bulk value due to the quantum confinement effect.  This finding shows importance
of the local environment on magnetic properties of the dopants in semiconductor
nanostructures.


\end{abstract}

\pacs{Valid PACS appear here}
\maketitle


\section{\label{sec:introduction}Introduction}

Further miniaturization of electronic devices implies addressing as
diminutive an amount of matter as possible. The ultimate goal of
modern semiconductor nano-electronics is the control and read-out of
single quantum states in solids~\cite{Awschalom2013, Kane1998}.
Optical orientation is a prospective method of addressing individual
quantum sized objects. Polarized photons possess a quantized amount
of spin angular momentum that transfers to charge carriers during
inter-band absorption process. Promising materials for optical
orientation applications are diluted magnetic semiconductors.
Photo-oriented carriers interact with magnetic impurities therefore
light can be used as a tool for the manipulation of spin states in
semiconductors. One of the most popular magnetic dopants in II/VI
and III/V materials is manganese \cite{Dietl2014}. The outer
electron shells of manganese consist of $4s^2$ and $3d^5$ orbitals,
in such a way that two s-electrons appear in the chemical bonds.
When manganese under this condition is placed in the cation
sublattice it behaves as an isoelectronic impurity in II/VI
semiconductors and acts as an acceptor in III/V materials. In II/VI
the spin of a charge carrier interacts with a large amount of
manganese spins forming a so called magnetic polaron. On the
contrary in III/V the manganese spin interacts with charge carriers
individually. The exchange interaction between the d-electrons and a
hole with the total momentum $J=3/2$ localized on the manganese acceptor
brings the neutral impurity to the ground state with full angular
momentum $F=1$~\cite{Karlik1982} while an ionized negatively charged
manganese acceptor has total spin $S=5/2$.

It is possible to control the spin of manganese ions and other
neutral acceptors in III/V based systems via optical pumping of the
band-to-band transition \cite{Myers2008, Leger2006, Kudelski2007,
Akimov2011}. However, when the ionized impurities absorb light with
the photon energy lower than the band gap~\cite{Eagles1960,
Dumke1963}, another process based on the impurity-to-band transition
becomes available in compensated semiconductors. It was recently
proposed to use this transition as another way to achieve a spin
polarization of the photoexcited electrons~\cite{Kokurin2013} or as
a method for read-out the Mn ion spin state~\cite{Govorov2004}. In
the present article we report observations of the optical
orientation of charge carriers in GaAs:Mn/AlGaAs quantum wells (QW)
achieved by impurity-to-band absorption.

The major advantage of using the impurity-to-band transition is a
potential possibility to address the impurities
individually~\cite{Koenraad2011}. Indeed, the optical orientation
via a band-to-band transition~\cite{Myers2008} deals with all the
impurities available within the carriers diffusion length. On the
contrary, the impurity-to-band transition deals only with ionized
impurities that were optically addressed.

Another advantage for using the impurity related optical transitions
is the absence of the Dyakonov-Perel spin relaxation
mechanism~\cite{Dyakonov1971, Dyakonov1972}. This mechanism is most
responsible for the spin relaxation of delocalized charge carriers
in GaAs, but it does not affect the localized impurity states. This
may lead to the significantly longer spin relaxation times of
photo-oriented carriers that are localized on impurities.

This article is organized as follows: the samples used in the present work are described in
Sec.~\ref{sec:samples}. In Sec.~\ref{sec:PL} the results of the photoluminescence (PL) characterization are
presented. There we also provide
evidence that the observed PL spectra involve a
recombination between an electron and the neutral Mn in the quantum well,
which we denote hereafter as $\mathrm{(e-Mn^0)}$
\[
\mathrm{e^- + Mn^0 \longrightarrow Mn^- + \hbar\omega}.
\]
The results of the PL experiments in an applied
magnetic field are presented in Sec.~\ref{sec:Zeeman}. The energy
position, a circular polarization of the luminescence in the
magnetic field and the Zeeman splitting are discussed in detail.
Section~\ref{sec:orientation} is devoted to the optical orientation
experiments carried out without magnetic field. We show that it is
possible to polarize the charge carrier spins in zero magnetic field
by optically pumping via the impurity-to-band transition. Finally,
Sec.~\ref{sec:conclusion} presents our conclusions.

\section{\label{sec:samples}Samples}
To address as small as possible amount of manganese spins during our
optical measurements, we have grown samples with narrow quantum
wells slightly doped with Mn. Each of the samples grown by molecular
beam epitaxy consist of a semi-insulating GaAs $[001]$-substrate
overgrown with a $1\,\mu$m buffer layer and the $3.7\,$nm GaAs
single quantum well sandwiched between $100\,$nm
Al$_{0.36}$Ga$_{0.64}$As barriers. In total six samples have been
grown: two undoped references and four delta-doped samples with the
manganese at a flux level of $10^{11}\,$cm$^{-2}$. The doped samples
differ in the position where Mn was introduced into the quantum
well. The delta-layer was inserted either at the interfaces, both
top and bottom of the well,  or directly into the quantum well, at
the 25\% or 75\% of the well width, respectively.

We obtained similar experimental results on each of the doped sample.
Most of the results that we report here were obtained on the
two samples that have been doped closer to the substrate.

\section{\label{sec:PL}Photoluminescence spectra and impurity-to-band excitation}

\begin{figure}[t]
\includegraphics[width=8cm]{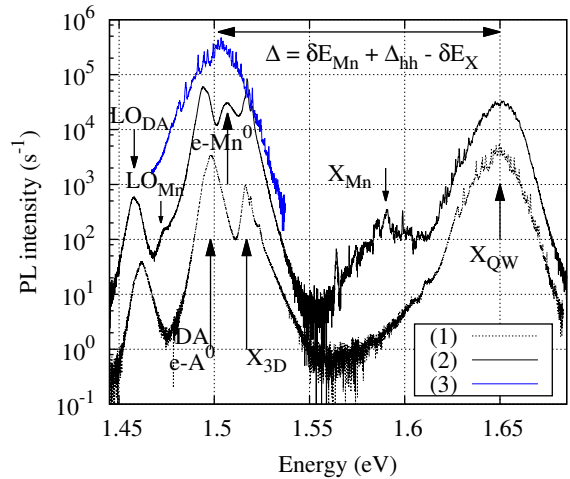}
\caption{\label{fig:PL} (Color online) Photoluminescence spectra of
the doped and undoped sample areas. The line (1) corresponds
to an undoped part of the sample. The spectrum contains a quantum
well (X$_{\rm QW}$) and the bulk (X$_{\rm 3D}$) excitons lines, the
usual donor-acceptor (DA) and the electron-acceptor
$\mathrm{(e-A^0)}$ transitions. The spectrum (2) displays
additionally an exciton localized on the Mn$^0$ acceptor (X$_{\rm
Mn}$) and an optical transition of the electron to the neutral Mn
$\mathrm{(e-Mn^0)}$. Phonon sidebands of the DA and the
$\mathrm{(e-Mn^0)}$ transitions are shown. The uppermost spectrum (3)
displays the quantum well related recombination that is shifted in
energy scale by $\Delta$ as indicated in the figure, see the text. }
\end{figure}

Figure~\ref{fig:PL} shows some typical PL spectra measured at 4$\,$K
and excited with the $632.8\,$nm--line of the HeNe laser. It turns
out that two distinctive types of the spectra can be detected from
our samples at these conditions. Most of the sample area shows a
spectrum typical for our undoped reference samples. This spectrum is
made up of three main components: a transition that corresponds to
recombination of the excitons localized inside the quantum well
(X$_{\rm QW}$), the usual lines of the  bulk exciton recombination
in the GaAs buffer layer (X$_{\rm 3D}$), and the radiative band of
the donor-acceptor transitions in the bulk material (DA). By
scanning the sample surface we have found small micron-sized spots
where two additional components become visible in the PL spectrum.
We assume that one of these lines is due to the quantum well
excitons localized on the manganese impurities X$_{\rm Mn}$ and the
other line corresponds to recombination of the electrons with the
neutral manganeses inside the quantum well, $\mathrm{(e-Mn^0)}$. As
it appears, the lateral distribution of Mn in our samples is
strongly non-uniform.

Due to the doping non-uniformity the luminescence properties related to Mn such as its transition energy or polarization
depend on the specific position of the photoexcited spot on the sample surface.
However, the particular properties that we have observed do reproduce themselves uniquely well
both in the Mn-poor and in the Mn-rich areas.

\begin{figure}[h]
\includegraphics[width=8cm]{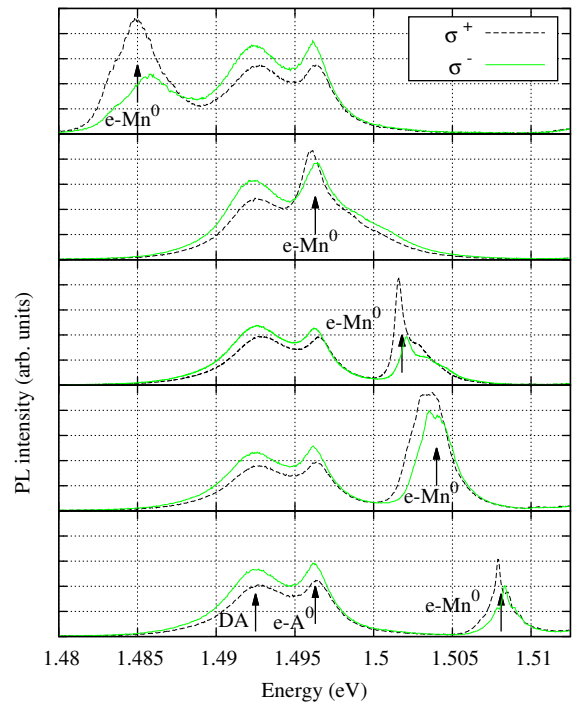}
\caption{\label{fig:polar} (Color online) Polarized
photoluminescence spectra in the applied magnetic field of 3$\,$T
under HeNe laser excitation taken at different spots of the sample
surface. The $\mathrm{(e-Mn^0)}$ transition energy fluctuates due to
local variations in the Mn$^0$ binding energy.}
\end{figure}

Let us first discuss the transition energy and the linewidth of the
$\mathrm{(e-Mn^0)}$ recombination. In our samples this transition
appears in the range 1.48--1.51$\,$eV. In the case of uniform doping
we expect to observe a band-to-impurity recombination line at the
energy lower than the band-to-band recombination and with
approximately the same linewidth as X$_{\rm QW}$. We illustrated
this hypothetical spectrum of the band-to-impurity transition with
the band-to-band recombination line shifted by energy $\Delta$, the
energy difference between the acceptor and the exciton related
transitions in the quantum well (Fig.~\ref{fig:PL}). Due to the
strong doping non-uniformity and the small amount of dopants we
observe not one unique broad luminescence band for
$\mathrm{(e-Mn^0)}$ transition but narrower bands randomly
positioned under the envelope luminescence of the X$_{\rm QW}$
downshifted in energy scale by $\Delta$ (see Fig. \ref{fig:polar}).
Figure \ref{fig:polar} depicts the polarized PL spectra measured at
different positions on the sample surface. The energy of the
band-to-impurity $\mathrm{(e-Mn^0)}$ luminescence strongly
fluctuates in the range $1.48 \div 1.51$eV due to lateral variations
in the quantum well width. An additional cause of the spectral
irregularity is a dependence of the Mn binding energy on its exact
position inside the quantum well. In such a narrow quantum well, the
dopants can take only a few discrete lattice positions. A difference
in the binding energy values of shallow impurities positioned in the
center of a quantum well or at the interface could reach
50\%~\cite{Bastard1981}. Our estimates for the deep impurities in
the zero--range potential approximation~\cite{Demkov1988} produce a
variation of the binding energy exceeding 10\%.

Controversial views about the energy $\Delta$ of the
$\mathrm{(e-Mn^0)}$ transition have been published so far
\cite{Bantien1987, Plot1986}. Some published results show that the
energy difference between the exciton and the acceptor related
transitions $\Delta$ approximately equals to the binding energy of
the acceptor and does not depend on the quantum well
width~\cite{Bantien1987}. On the other hand, manganese is considered
to be a deep acceptor in GaAs, and its binding energy should not be
strongly affected by the quantum confinement. Then the energy
difference between the exciton-- and the acceptor--related
transitions is given by $\Delta = (\delta E_{\rm
Mn}+\Delta_{hh}-\delta E_X)$~\cite{Plot1986}, where $\delta E_{\rm
Mn}\,$ is the acceptor binding energy, $\Delta_{hh}$ is the quantum
confinement of the heavy--hole subband, $\delta E_X$ is the exciton
binding energy. It should be pointed out that the observed energy
difference strongly decreases with an increasing acceptor
concentration inside the quantum wells~\cite{Holtz1998}. Because in
this paper we are dealing with the samples of low manganese
concentration, the observed $\mathrm{(e-Mn^0)}$ transition energy
includes the hole confinement shift, in agreement with results
published in~\cite{Plot1986}. We obtain $\Delta_{hh}=29\,$meV using
a numerical solution for the well width $L =3.7\,$nm with the
barriers height 200$\,$meV,  and the  heavy hole mass $m_{hh} =
0.51$ in the well and $m_{hh}=0.60$ in the barriers. The exciton
binding energy is taken as $\delta E_X \sim 11\,$meV
\cite{Mathieu1992}. We use the manganese binding energy $\delta
E_{\rm Mn} = 130\,$meV, which is increased from its bulk value
$\delta E_{\rm Mn} = 113\,$meV due to the effect of quantum
confinement \cite{Romanov2012}.

Dopants in semiconductors are always partly ionized due to technologically unavoidable small amount of doping compensation.
In our samples the degree of manganese ionization is enhanced
because the quantum well is subjected to electric fields of the surface depletion layer.
This makes it possible to optically pump the photoluminescence using an impurity related absorption.
In such case an ionized manganese acceptor captures the pump photon of energy less than the band gap in the quantum well.  That produces
a photoexcited electron in the quantum well and the neutral acceptor.

To optically pump the samples, our setup is equipped with two lasers. A HeNe
laser pumps the quantum well related luminescence and we employ
a Ti-sapphire laser tuned to 1.53 eV in order to pump the
impurity-to-band transition.
In a two-color experiment the both lasers are focused on the same $\mu$PL spot.
In our measurements under the Ti-sapphire laser excitation, the
$\mathrm{(e-Mn^0)}$ line appears only at the spots where both
X$_{\rm Mn}$ and $\mathrm{(e-Mn^0)}$ transitions can be detected under the HeNe excitation first.
This observation is the direct
evidence that here we deal with the impurity-to-band excitation inside the
GaAs/AlGaAs quantum well doped with partially compensated impurities.

\section{\label{sec:Zeeman}Polarized photoluminescence in magnetic field and Zeeman splitting}

\begin{figure}[b]
\includegraphics[width=8cm]{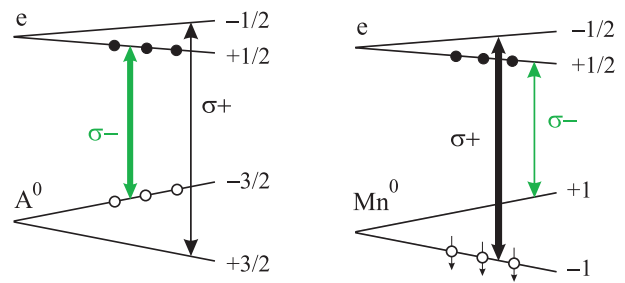}
\caption{\label{fig:zeeman1} (Color online) Recombination schemes
for the shallow acceptor (left) and Mn (right). The ground state in
magnetic field is denoted with black circles. The thicker vertical arrows correspond to higher transition probabilities.}
\end{figure}

The photoluminescence of $\mathrm{(e-Mn^0)}$ recombination in GaAs
has an opposite sign of the circular polarization in comparison with
a shallow acceptor in the magnetic field~\cite{Karlik1982}. This is
caused by the exchange interaction between the localized hole and
and the $3d^5$ core electrons of Mn. The Hamiltonian of a manganese
acceptor in the applied magnetic field $\bm{B}$,  including the
exchange interaction,  is given by
\begin{equation}
H_0 = -A\,\bm{J\cdot S} + g_s \mu_B \bm{S\cdot B} + g_h \mu_B
\bm{J\cdot B}.
\end{equation}
Here $A<0$ is the exchange parameter,  $\bm{S}$ and $\bm{J}$ are the
spin of the d-electrons and the hole angular momentum, respectively,
with $g_s$ and $g_h$ the corresponding g-factors, and $\mu_B =
|e|\hbar/2mc$ is the Bohr magneton.

It is easy to find all 12 possible optical
$\mathrm{(e-Mn^0)}$ transition probabilities using
 the selection rules in dipole approximation and taking into account the angular
momentum coupling in the ground state
$F=1$ (see table~\ref{fig:table1}).

\begin{table}[h]
\begin{center}
\begin{ruledtabular}
\begin{tabular}{c c c c c c c }
$F_z$, $S_z$ & +5/2 & +3/2 & +1/2 & -1/2 & -3/2 & -5/2 \\
\hline
 +1 & $10\sigma^-\uparrow$ & $2\sigma^-\downarrow$ & $1\sigma^+\uparrow$ & $1\sigma^+\downarrow$ & 0 & 0 \\
 0 & 0 & $4\sigma^-\uparrow$ & $2\sigma^-\downarrow$ & $2\sigma^+\uparrow$ & $4\sigma^+\downarrow$ & 0\\
 -1 & 0 & 0 & $1\sigma^-\uparrow$ & $1\sigma^-\downarrow$ & $2\sigma^+\uparrow$ & $10\sigma^+\downarrow$\\
\end{tabular}
\end{ruledtabular}
\end{center}
\caption{\label{fig:table1} All
possible optical $\mathrm{(e-Mn^0)}$ transitions.
Projections of the electron spin are indicated with arrows.}
\end{table}

In the narrow quantum wells, the quantum-confined splitting $\delta$
between the acceptor states $J_z = \pm 3/2$ and $J_z = \pm 1/2$ can
be comparable in value with the exchange parameter $|A|$, where
$|A|$ is about $1\div 5\,$meV \cite{Nestoklon2015}. For a shallow
acceptor in 3.7-nm thick ${\rm GaAs/Al_{0.3}Ga_{0.7}As}$ quantum
well parameter $\delta$ is about $10\div 12$ meV
\cite{Masselink1985}, and for a deep center this value can be
further reduced. Thus, we are likely to find ourselves in the
intermediate situation, when $|\delta|\gtrsim |A|$. In order to
describe the quantum--well confinement we use an extra term in the
Hamiltonian:
\begin{equation}
H = \frac{\delta}{2}\left(J^2_z - \frac{5}{4}\right).
\end{equation}

Note, that in the 2D-case,  when $|\delta|\gg|A|$, the ground--state
wavefunctions are composed of $\Psi^S_{\pm 5/2}\Psi^J_{\mp 3/2}$
components exclusively. In the 3D--limit, when $|\delta|\ll|A|$, the
wavefunctions remain dominated by the same heavy-hole components as
before, although the ground state includes an admixture of the light
holes~\cite{Averkiev1988}. However, there will be no qualitative
difference in the main character of the ground states in these two
limiting cases. In the 2D-description, which we use from now on,
there are only two possible optical transitions given by:
\begin{equation}
\label{twostates}
\begin{split}
-1/2+(-5/2;+3/2)\rightarrow (-5/2)+\sigma^+,\\
+1/2+(+5/2;-3/2)\rightarrow (+5/2)+\sigma^-.
\end{split}
\end{equation}

\begin{figure}[t]
\includegraphics[width=8cm]{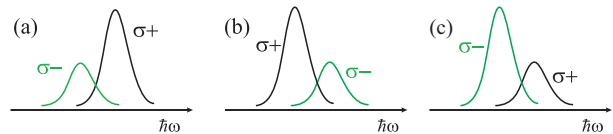}
\caption{\label{fig:zeeman2} (Color online) Schematic outline of the
polarized photoluminescence spectra for optical transitions of the
electron to: (a) Mn in bulk GaAs; (b) Mn in a quantum well and (c) a shallow
acceptor in bulk GaAs.}
\end{figure}

\begin{figure}[t]
\includegraphics[width=8cm]{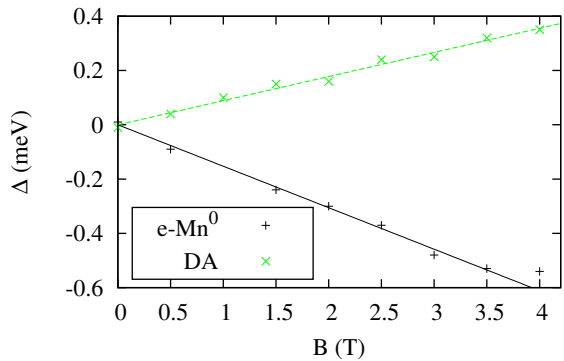}
\caption{\label{fig:split} (Color online) Zeeman splitting
(experimental points and a linear fit) of DA-transition occurring in
the bulk GaAs substrate region and $\mathrm{(e-Mn^0)}$ transition in
the quantum well. The experimental points are from the spectra in
Fig.~\ref{fig:pollum} for the maximums of $\sigma^{\pm}$ peaks.}
\end{figure}

The left panel in Fig.~\ref{fig:zeeman1} sketches the optical
transitions of a conduction electron to a common shallow acceptor.
In the applied magnetic field the ground state $J_z = -3/2$  is
populated with a hole. The $\sigma^-$ component in this case
dominates in the circular polarized photoluminescence. As shown in
Fig.~\ref{fig:zeeman2}~(c) the dominating $\sigma^-$ line is also
shifted to the lower energies. This scheme is in full agreement with
our measurements as well as with the results published
before~\cite{Bimberg1978}. The Zeeman splitting for the shallow
acceptor in our measurements (Fig.~\ref{fig:split}) produces
$g_h=+0.36 \pm 0.01$, assuming the electron g-factor bulk value of
$-0.44$.

The right panel in Fig.~\ref{fig:zeeman1} presents the optical
transitions involving Mn. Here the situation is more complex due to
the Zeeman splitting of the final state after recombination. The
initial state of Mn is $F_z = -1$ ($-5/2;+3/2$) as discussed before.
In this case the $\sigma^+$-polarized transition dominates in the
Mn-related photoluminescence, see Eq.(\ref{twostates}). This
analysis is supported by the experimental results firstly published
in~\cite{Karlik1982}

In Fig.~\ref{fig:polar} we have shown the polarized PL spectra
measured at different spots of the doped samples. The well-resolved
$\mathrm{DA}$ and $\mathrm{(e-A^0)}$ transitions at 1.492$\,$eV and
1.497$\,$eV clearly show a negative sign of the polarization. As
discussed above, the energy of $\mathrm{(e-Mn^0)}$ transition varies
across the sample. However, in all PL spectra the
$\mathrm{(e-Mn^0)}$ transition has a positive sign of the
polarization, independent on the spot where the photoluminescence is
collected. This observation supports our assignment of the
transitions to the Mn inside quantum well.

Now let us discuss the Zeeman splitting of the $\mathrm{(e-Mn^0)}$
transition. Here we assume that the Zeeman splitting is much less
than both the exchange and the quantum-confined splittings. In bulk
GaAs the complexity of this transition is due to splitting of the
final state. This splitting does not affect the PL polarization
degree but it has to be accounted in the spectral position of the
polarized components. Here we neglect the splitting in the
conduction band because the g-factor of electrons is small in
comparison with the g-factor of holes in such quantum
wells~\cite{Hannak1995}. Then we can express the splitting $\Delta
E$ of the $\mathrm{(e-Mn^0)}$ transition as $\Delta E = (5g_s -
2g_F) \mu_B B$. Using $g_F=2.77$ and $g_s=2$ from the ESR
measurements~\cite{Schneider1987} we get a positive splitting and
the dominating $\sigma^+$ component at a higher energy than the
$\sigma^-$ component as shown in Fig.~\ref{fig:zeeman2}~(a). That
kind of splitting was experimentally observed in bulk
GaAs:Mn~\cite{Kikkawa1994}. Our experiments also produce the
dominating $\sigma^+$ component but at the lower energy as
illustrated in Fig.~\ref{fig:zeeman2}~(b). For the reference, we
also depict in the same figure (see Fig.~\ref{fig:zeeman2}~(c)) a
shallow acceptor in the bulk GaAs with a sign assignment of the
splitting from \cite{Bimberg1978}.

\begin{figure}[b]
\includegraphics[width=8cm]{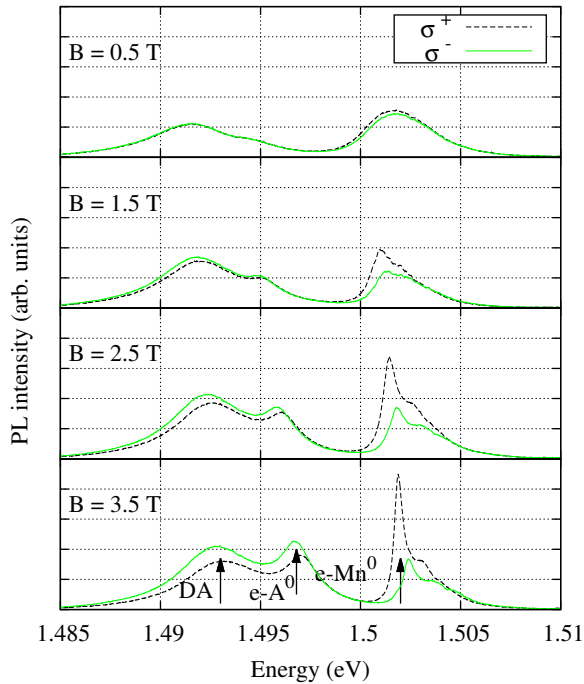}
\caption{\label{fig:pollum} (Color online) Polarized
photoluminescence spectra of bulk DA, $\mathrm{(e-A^0)}$ and
$\mathrm{(e-Mn^0)}$ transitions taken at different magnetic fields.}
\end{figure}

The most probable reason for the observed sign of
$\mathrm{(e-Mn^0)}$ Zeeman splitting is a reduction of the hole
g-factor due to the effect of strong quantum confinement.  The
g-factor of the core d-electrons is independent on material
parameters because of their atomic-scale confinement. On the
contrary, the $\Gamma_8$ hole g-factor, in addition to a
paramagnetic contribution described by the Luttinger material
parameter $\kappa$ (see, for instance~\cite{Luttinger1956}), has
also an orbital contribution~\cite{Bir1967}. And the quantum
confinement considerably changes the hole
g-factor~\cite{Wimbauer1994, Bree2012}. A hole bound by an acceptor
already displays a renormalized g-factor even in the bulk
semiconductor ~\cite{Gelmont1973}. A g-factor of the acceptor-bound
hole in the quantum well has substantially different value (possibly
even an opposite sign) depending on a ratio between the Bohr radius
and the quantum well width. Numerous earlier experiments
demonstrated that the hole g-factor is strongly modified by the size
and geometrical anisotropy of the nanostructure,  and also depends
on a given position of the impurity atom inside the
nanostructure~\cite{Bree2012, Snelling1992, Krebs2009,
Kudelski2007}.

In bulk GaAs the Mn g-factor in the ground state $F=1$ can be
expressed as $g_F = 7/4g_s - 3/4g_h$ with
$g_h=+1$~\cite{Averkiev1988}. In narrow quantum wells the energy
structure of Mn undergoes a change and, in the 2D limit, can be
described with only two states  $\Psi^S_{-5/2}\Psi^J_{+3/2}$ and
$\Psi^S_{+5/2}\Psi^J_{-3/2}$ corresponding to $F_z=\pm 1$ as already
expressed in Eq.~(\ref{twostates}) with the g-factor $g_F = 5/2g_s -
3/2g_h$. Hence in the 2D-model the spectral splitting in magnetic
field depends on the hole g-factor only, $\Delta E=3g_h\mu_BB$ as
the Zeeman splitting of the d-shell electrons in initial and final
states cancels out. By taking into account the experimentally
measured value of splitting presented in Fig.~\ref{fig:split} and
Fig.~\ref{fig:pollum}, we obtain in the 2D limit $g_h = -0.86\pm
0.06$. The actual values of the Mn hole g-factor vary in the range
$g_h = -1.5 \div -0.5$ at different positions on the sample
(Fig.~\ref{fig:polar}). Such a strong variation of the g-factor is
caused by fluctuations in the quantum well width and by different
actual positions of the impurity inside the quantum well. This
result is in a good agreement with recent measurements of the
Mn--related PL in InAs quantum dots~\cite{Krebs2013} where authors
also observed strong fluctuations in the Mn-related g-factor.

To conclude this section, we would also briefly mention that our
2D-model produces the value of $g_h$ of about $-1$ that is
renormalized from the bulk value of $+1$ by the quantum confinement,
while the 3D-limit would result in a value down to $-4$.

\section{\label{sec:orientation}Optical orientation in zero magnetic field}

\begin{figure}[t]
\includegraphics[width=8cm]{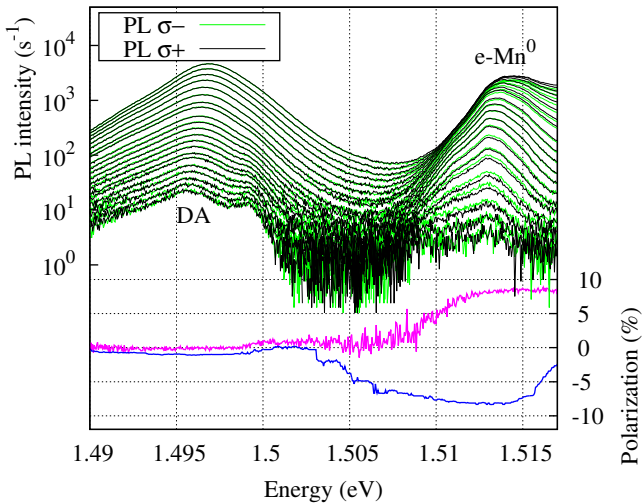}
\caption{\label{fig:int} (Color online) Polarized photoluminescence
spectrum of the bulk DA-line and $\mathrm{(e-Mn^0)}$ transition
under $\sigma^+$ excitation $\hbar\omega \approx 1.53\,$eV measured
at different pump power $I = 0.1\div30\,\mu$W in zero magnetic
field. The upper (purple) and the lower (blue) dependencies of the
polarization degree on photon energy were measured at 30 and
0.6$\,\mu$W pump power, respectively. The lowermost curve is
smoothed for clearness.}
\end{figure}

Despite the fact that the optical orientation in semiconductors is a
well-studied phenomenon~\cite{Meier1984}, so far as we know, nobody
used the impurity-to-band transition to optically orient the charge
carriers spins. We carry out our experiments in a traditional
optical orientation framework exciting impurity-to-band transition
with a $\sigma^+$~circularly polarized light and detecting the
polarized photoluminescence. Our results on the optical orientation
measurements significantly vary from point to point on the sample
similarly to the above described magneto-luminescence experiments.
Here we describe the three most representative properties of the
measured spectra: 1) a higher-energy broadening of the spectra at
high pump power, 2) the spectral dependence of polarization and 3)
the dependence of polarized photoluminescence intensity on the pump
density. Our aim is to show that all these properties can be
described within one consistent model.

Let us consider a limited amount of partly compensated manganese acceptors inside the quantum well.
A binding energy  of the acceptor is fluctuating due to a randomized distance between that acceptor and the barriers.
Acceptor states at the lower energies remain always neutral while the acceptors at the higher energies are ionized
due to the doping compensation. We shall refer to these neutral acceptors as equilibrium Mn$^0$.
Under intensive photoexcitation via the impurity-to-band transition, the ionized acceptors should also become neutral.
In the same way, we shall refer to these acceptors as non-equilibrium Mn$^0$ acceptor states.

Figure~\ref{fig:int} shows the polarized PL spectra taken at
different pump power in zero magnetic field while the pump photon
energy was set to $\hbar\omega \approx 1.53\,$eV. We observe a
strong high-energy broadening of the $\mathrm{(e-Mn^0)}$ line while
the DA line spectral position and its spectral shape is not changed.
This high-energy broadening is caused by the photo-induced
neutralization of the high-energy acceptor states leading to a
production  of the non-equilibrium Mn$^0$.

\begin{figure}[h]
\includegraphics[width=8cm]{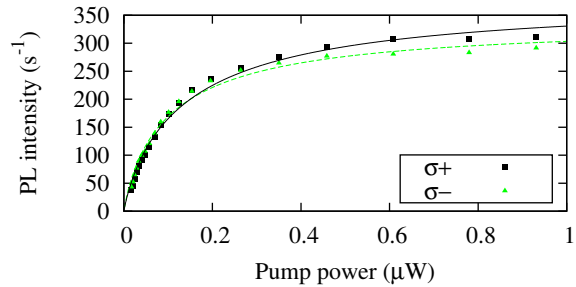}
\caption{\label{fig:int3} (Color online) Polarized
$\mathrm{(e-Mn^0)}$ photoluminescence vs pump power. The
experimental points (squares and triangles) are fitted by Eq.~(4)
(solid and dashed lines).}
\end{figure}

The intensity of $(\mathrm{e-Mn^0})$ transition versus pump power
(Fig.~\ref{fig:int3}) shows a dependence characteristic for the
impurity--related luminescence. It is linear in limit of the low
pump power and saturates at the higher pump power. We fit this
experimentally observed dependence $\mathcal{I}(G)$ by a
two-parameter function:

\begin{equation}
\label{fittingf}
\mathcal{I}(G)=\frac{\beta^2G^2}{4\mathcal{I}_S}\left(\sqrt{1+\frac{4\mathcal{I}_S}{\beta
G}}-1\right)^2,
\end{equation}
where $G$ is the pump power, $\mathcal{I}_S$ is a value of the
saturated intensity, and $\beta$ is a parameter that is set equal to
a slope of the experimentally observed dependence $\mathcal{I}(G)$
at zero pump power. The fitting function $\mathcal{I}(G)$ is easily
derived for the case of nearly total doping compensation from the
following rate equation:

\begin{equation}
\label{kinetics} \frac{dn}{dt}=m^-\alpha^{imp} VG-\gamma nm^0,
\end{equation}
here $n$ is a concentration of the photo-excited electrons, and
$m^-$ and $m_0$ are the concentrations of ionized and neutral Mn
acceptors, respectively; $\alpha^{imp}$ is the ionized impurity
absorption coefficient and $\gamma$ is the coefficient of
bimolecular recombination; $V$ is the photoexcitation volume, which
is equal to a product of the laser spot area and the width of the
quantum well, $V\sim 3 \cdot 10^{-15}$ cm$^{3}$. To solve
Eq.~(\ref{kinetics}), one has to include charge conservation $m^-+
m_0=M$ and $m_0=n$, where $M$ is the total concentration of the Mn
dopants.

It follows then that the concentration $M$ can be estimated as:

\begin{equation}
\label{avogadro} M\sim\frac{\mathcal{I}_S\alpha^{imp}
V}{\beta\gamma}.
\end{equation}

Substituting $\alpha^{imp}\sim 100$ cm$^{-1}$, and $\gamma=10^{-10}$
cm$^3\cdot$s$^{-1}$ in Eq.~(\ref{avogadro}) with experimental values
$\mathcal{I}_S=300$ s$^{-1}$ and $\beta=8.4\cdot10^{-18}$ cm$^2$
taken in units of the pump light flux, we estimate the total Mn
concentration as $M\sim 10^{17}$ cm$^{-3}$ that corresponds to the
surface density of about $10^{11}$ cm$^{-2}$ or  
to a few hundred Mn atoms within the excitation
area in a diffraction limit.

\begin{figure}[t]
\includegraphics[width=8cm]{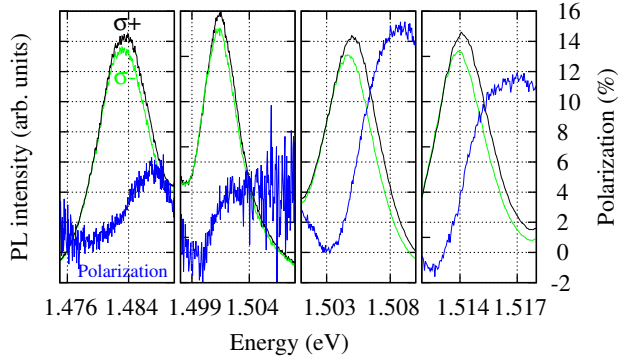}
\caption{\label{fig:int2} (Color online) Polarized photoluminescence
spectra of $\mathrm{(e-Mn^0)}$ transition under $\sigma^+$
excitation measured in different spots on the sample in zero
magnetic field.}
\end{figure}

The polarization degree and the photoluminescence intensity
saturates as the pump power is increased. The value of maximum
polarization at the saturation point fluctuates from spot to spot in
the range 5--15$\,$\% for the pump photon energy $\hbar\omega
\approx 1.53\,$eV. It becomes higher if the transition energy is
closer to the pump energy, which is a typical behavior for the
optical orientation. In the majority of the studied spots the
polarization of photoluminescence has a distinctive spectral
dependence (Fig.~\ref{fig:int2}). Under the $\sigma^+$-pump the
electron-to-manganese PL is mostly $\sigma^+$-polarized at the
high-energy side of spectrum while the polarization is close to zero
at the low-energy side. We suppose that the specific polarization of
manganese acceptors affects the polarized spectra: the
non-equilibrium high-energy acceptor states are polarized via
impurity-to-band transition while the equilibrium acceptors keep
zero net-polarization.

\begin{figure}[b]
\includegraphics[width=8cm]{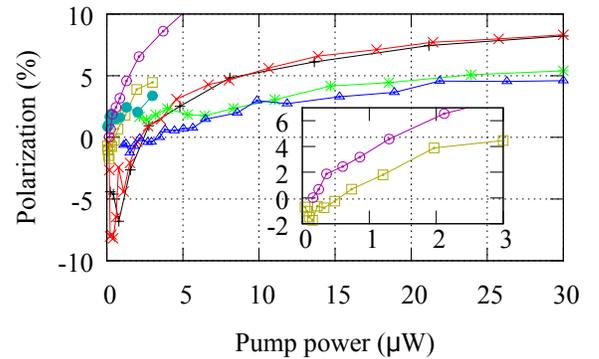}
\caption{\label{fig:int4} (Color online) Dependence of
$\mathrm{(e-Mn^0)}$ photoluminescence polarization on pump power
measured at different spots on the sample under the pump with
787$\,$nm wavelength. Inset shows two curves measured at the same
spot but with different pump wavelength: circles --- 810$\,$nm,
squares --- 788$\,$nm.}
\end{figure}

Fig.~\ref{fig:int4} depicts the dependence of PL polarization degree
vs pump power. Here we observe an intriguing phenomenon: the sign of
polarization is changed from negative to positive while pump power
increases (Fig.~\ref{fig:int}). It is a quite rarely observed
phenomenon and its description has to employ a complicated scheme of
spin and energy relaxation processes~\cite{Shabaev2009}. Here we
propose an approximate description of the polarization sign
inversion in our experiments based on the suggested model of the
impurity-to-band transition.

\begin{figure}[t]
\includegraphics[width=8cm]{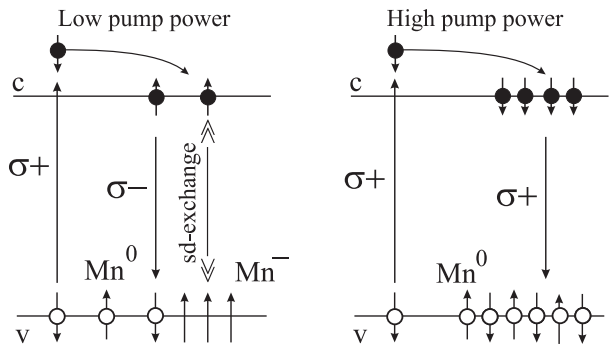}
\caption{\label{fig11}Sign reversal mechanism. At low pump power the
negative polarization degree of the photoluminescence is due to an
exchange interaction of optically oriented Mn$^-$ and photoexcited
electrons. At high pump power the photoexcited electrons keep its
spin and provide the positive polarization sign.}
\end{figure}

The positive polarization, as we observe in the limit of high pump
power, corresponds to an ordinary optical spin orientation of the
photoexcited electrons. In this regime, the spin relaxation time can
be estimated from the degree of polarization. Indeed, the lifetime
of the bimolecular recombination equals $\tau_l=(\gamma n)^{-1}$ and
from Eq.~(\ref{kinetics}) one can easily derive
\begin{equation}
\label{lifetime} \tau_l^{-1}=\frac{\gamma
M}{2}\left(\frac{G}{G_0}\right)\left(\sqrt{1+\frac{4G_0}{G}}-1\right),
\end{equation}
where we introduced characteristic pump power $G_0=\frac{\gamma
M}{\alpha^{imp} V }$. In the limit of high pump power ($G\gg  G_0$),
when the $(\mathrm{e-Mn^0})$ transition is saturated, the radiative
lifetime $\tau_l$ becomes constant $\tau_l^{\infty}=(\gamma
M)^{-1}$. In other words,  in this limit
there are
many photo-charge carriers generated within the photo absorption
(photo-neutralization) cross section of $\mathrm{Mn}^-$, that is $G
\tau_l^{\infty} \gg (\alpha^{imp} V)^{-1}$. In our case
this leads to an estimate of $\tau_l^{\infty} \lesssim 100$\,ns, the
value that is in a good agreement with the radiative lifetimes in
compensated GaAs:Mn reported elsewhere \cite{Astakhov2008}. The
corresponding electron spin relaxation time $\tau_s$ can be then
estimated from the experimental polarization values following the
dependence of the polarization degree $P$ on the ratio between
$\tau_l$ and $\tau_s$~\cite{Meier1984}:
\begin{equation}
\label{orientation}
P=\frac{P_0}{1+\tau_l/\tau_s}.
\end{equation}
The maximum polarization value of $P_0=100\%$ can only be achieved
in the 2D-case if $\tau_l \ll \tau_s$. Then the photo-excited
electrons keep their spins oriented until they recombine, with the
PL polarization positive due to the selection rules
(Fig.~\ref{fig11}, right panel). In our case, the
$\mathrm{(e-Mn^0)}$ transition displays polarization of 5--15\% that
translates into $\tau_s \approx 10$\,ns, a typical value for this
material. We note that for the optically saturated
$\mathrm{(e-Mn^0)}$ transition the magnetic nature of impurity does
not play any role in the optical orientation of the spins. Identical
behavior would be equally observed in the case of a non-magnetic
dopant.

This is different from when the transition is not saturated, in the
limit of low pump power $G\ll G_0$, and here we infer that the
negative polarization is due to the net spin polarization of the
ionized $\mathrm{Mn}^-$. Indeed, in this regime the radiative
recombination slows down as $G$  decreases and the radiative times
become dependent on $G$ itself:
$\tau_l^{(0)}=\tau_l^{\infty}({G_0}/{G})^{^1/_2}$. In our
experiments this regime is realized at $G$ of a few tenth of $\mu$W.
From Eq.~(\ref{lifetime}), this pump power corresponds to a
microsecond range of the bimolecular recombination. That closely
matches the spin relaxation time reported for the manganese ions in
GaAs~\cite{Akimov2011}. In this limit the optical spin orientation
of the photoexcited electrons does not contribute to the PL
polarization as $\tau_l\gg\tau_s$ now. It is now the d-electrons of
ionized manganese Mn$^-$ that become optically oriented
(Fig.~\ref{fig11}, left panel). In accordance with
Eq.~(\ref{twostates}), only $\mathrm{Mn^-}$ with $S_z=-5/2$ absorbs
the $\sigma^+$ photons. Then the photo-excited $\mathrm{Mn^0}$
acceptors and electrons lose their spin shortly after the
photoexcitation. Once they recombine, an equal amount of
$\mathrm{Mn^-}$ with $S_z=-5/2$ and $S_z=+5/2$ gets back into the
system, that is to say that the $\mathrm{Mn^-}$ ions with $S_z=+5/2$
are accumulated. Despite the fact that $\tau_l \gg \tau_s$ the
photo-excited electrons become spin-polarized via the exchange
interaction with the neighboring $\mathrm{Mn}^-$ in the predominant
$S_z=+5/2$ state. This provides a negative sign of the polarization
according to Eq.~(\ref{twostates}).

The optical orientation measurements carried out at the different
pump wavelengths (see inset of Fig.~\ref{fig:int4}) support our
description. Under the pump with higher photon energy the electron
polarization becomes lower due to stronger spin relaxation of
photo-excited electrons. In that case the PL polarization is
determined by polarization of Mn$^-$ ions and gets the negative sign
at low pump power. Under the pump with lower photon energy the
photo-excited electrons conserve its spins and PL polarization is
positive.

\section{\label{sec:conclusion}Conclusion}

We carried out magneto-optical and optical orientation experiments
using the impurity-to-band excitation in Mn doped narrow quantum
wells. The g-factor of hole localized on the Mn acceptor was found
to be modified due to the quantum confinement effect from its bulk
value $g_{h} = +1$ to $g_{h} = -0.5 \div -1.5$. We have shown that
it is possible to optically polarize the charge carriers by means of
impurity-to-band excitation. We also observe a sign inversion in the
circular polarization of the Mn-related luminescence and suggest a
model based on the impurity-to-band excitation to explain this
experimental observation by optical orientation of the Mn ions.

\begin{acknowledgments}
We acknowledge funding from Russian Science Foundation. P.V.P.,
N.S.A., P.M.K. and A.Yu.S. were supported by project 14-42-00015
(experiments and general discussion). I.A.K. and Yu.L.I. were
supported by project 14-12-00255 (theory).
\end{acknowledgments}


%

\end{document}